\documentclass[a4paper,english,final,1p,times]{elsarticle}
\usepackage[T1]{fontenc}
\usepackage[latin9]{inputenc}
\usepackage{color}
\usepackage{babel}
\usepackage{units}
\usepackage{graphicx}
\usepackage[unicode=true,
 bookmarks=true,bookmarksnumbered=true,bookmarksopen=true,bookmarksopenlevel=2,
 breaklinks=false,pdfborder={0 0 1},backref=false,colorlinks=true]
 {hyperref}
\hypersetup{pdftitle={Introduction to LyX},
 pdfauthor={LyX Team},
 pdfsubject={LyX-documentation Intro},
 pdfkeywords={LyX, documentation},
 linkcolor=black, citecolor=black, urlcolor=blue, filecolor=blue, pdfpagelayout=OneColumn, pdfnewwindow=true, pdfstartview=XYZ, plainpages=false}

\makeatletter

\pdfpageheight\paperheight
\pdfpagewidth\paperwidth

\usepackage{ifpdf} 
\ifpdf 

 \IfFileExists{lmodern.sty}{\usepackage{lmodern}}{}

\fi 

\let\myTOC\tableofcontents
\renewcommand\tableofcontents{%
  \pdfbookmark[1]{\contentsname}{}
  \myTOC
  \cleardoublepage
  \pagenumbering{arabic} }

\def\LyX{\texorpdfstring{%
  L\kern-.1667em\lower.25em\hbox{Y}\kern-.125emX\@}
  {LyX}}

\journal{Nuclear Physics A}

\makeatother

\begin{document}

\title{Beam energy and centrality dependence of the statistical moments
of the net-charge and net-kaon multiplicity distributions in Au+Au
collisions at STAR}

\author{Daniel McDonald (for the STAR Collaboration)}

\address{Rice University, 6100 Main Street, Houston, TX 77251}
\begin{abstract}
In part to search for a possible critical point (CP) in the phase
diagram of hot nuclear matter, a Beam Energy Scan was performed at
the Relativistic Heavy-Ion Collider at Brookhaven National Laboratory.
The STAR experiment collected significant Au+Au data sets at beam
energies, $\sqrt{{\rm s}_{\rm NN}}$, of 7.7, 11.5, 19.6, 27, 39,
62.4, and 200 GeV. Lattice and phenomenological calculations suggest
that the presence of a CP might result in divergences of the thermodynamic
susceptibilities and correlation length. The statistical moments of
the multiplicity distributions of particles reflecting conserved quantities,
such as net-charge and net-strangeness, are expected to depend sensitively
on these correlation lengths, making them attractive tools in the
search for a possible critical point. The centrality and beam-energy
dependence of the statistical moments of the net-charge multiplicity
distributions will be discussed. The observables studied include the
lowest four statistical moments (mean, variance, skewness, kurtosis)
and the products of these moments. The measured moments of the net-kaon
multiplicity distributions will also be presented. These will be compared
to the predictions from approaches lacking critical behavior, such
as the Hadron Resonance Gas model and Poisson statistics.
\end{abstract}
\maketitle

\section{Introduction}

The phase diagram of nuclear matter, ${\it i.e.}$, the temperature
versus the baryochemical potential, is largely speculative. There
is indirect evidence in Lattice Quantum Chromodynamics (QCD) of a
first-order phase transition at the boundary of the Quark-Gluon Plasma
(QGP) and a hadron gas at higher values of the baryochemical potential
\citep{Aoki06}. At the top energy of the Relativistic Heavy-Ion Collider
(RHIC), $\sqrt{{\rm s}_{\rm NN}}$=200 GeV, the transition between
the QGP and a hadron gas is a crossover \citep{STAR}. Thus, a critical
point (CP) could exist. In order to explore the QCD phase diagram
and locate the possible critical point, a Beam Energy Scan (BES) was
performed at RHIC in 2010 and 2011. Data at seven different beam energies
between 7.7 and 200 GeV were collected by the Solenoidal Tracker at
RHIC (STAR) experiment, corresponding to the range of baryochemical
potential of \textasciitilde{}20 to \textasciitilde{}420 MeV \citep{Cle06}.

The characteristic signatures of a CP are the divergences of the correlation
length, $\xi$, and the susceptibilities of conserved quantities,
such as net-strangeness and net-charge \citep{Cheng09,KR11}. These
divergences should be reflected in the moments products $S\sigma$
and $\kappa$$\sigma^{2}$, where $\sigma$ is the standard deviation,
$S$ is the skewness, and $\kappa$ is the kurtosis. The higher the
involved moment, the larger the CP-related power of the divergence
with the correlation length: $S\sigma$\textasciitilde{}$\xi^{2.5}$
and $\kappa$$\sigma^{2}$\textasciitilde{}$\xi^{5}$ \citep{Steph09}.

Various theoretical approaches predict significant enhancements of
the moments products near the CP. In a Polyakov loop extended quark-meson
model, a \textasciitilde{}50\% enhancement of $\kappa$$\sigma^{2}$
is predicted \citep{Sko12}. In Ref. \citep{MIT}, an ansatz for the
dependence of the critical component of the fluctuations on the baryochemical
potential was made based on universality arguments with parameters
suggested by lattice calculations. This results in a critical contribution
that is a peak with a width of $\sim$100 MeV. This critical component
is predicted to result in a 1-2 order-of-magnitude enhancement of
the kurtosis of pions and protons, which make up a majority of the
measured charge. Different average values of baryochemical potential
are explored in collisions with different $\sqrt{{\rm s}_{\rm NN}}$
values \citep{Cle06}. Thus, the expected enhancement should be $\sqrt{{\rm s}_{\rm NN}}$-localized.
The expectation is a non-monotonic behavior of the moments products
for nuclear systems that freeze out close to the CP.

\section{Analysis Details and Results}

Particle identification was done with a combination of the ionization
energy loss in the Time Projection Chamber (TPC) and the information
from the Time-of-Flight (TOF) detector \citep{TOF}. Event-by-event
identified particle multiplicities were measured in STAR's full azimuthal
coverage and within $\pm$0.5 units of psuedorapidity ($\eta$). Mathematical
autocorrelations were avoided by defining the centrality using the
number of charged tracks with 0.5$<$$|\eta|$$<$1.0. Analysis was
done for all charged tracks with transverse momenta from 0.2<$p_{\rm{T}}$<2.0
GeV/$c$. A centrality bin-width correction was done to avoid fluctuations
resulting from the centrality width of the analysis bins \citep{Luo11}. 

While the moments products should be very sensitive to possible critical
fluctuations, they could also be very sensitive to experimental sources
of fluctuations. Quality assurance was performed to remove aberrant
runs at each beam energy. In addition, careful studies were performed
to investigate and remove bad events from the data sets. 

\begin{figure}
\raggedright{}%
\begin{minipage}[t]{0.5\columnwidth}%
\includegraphics[width=1\columnwidth]{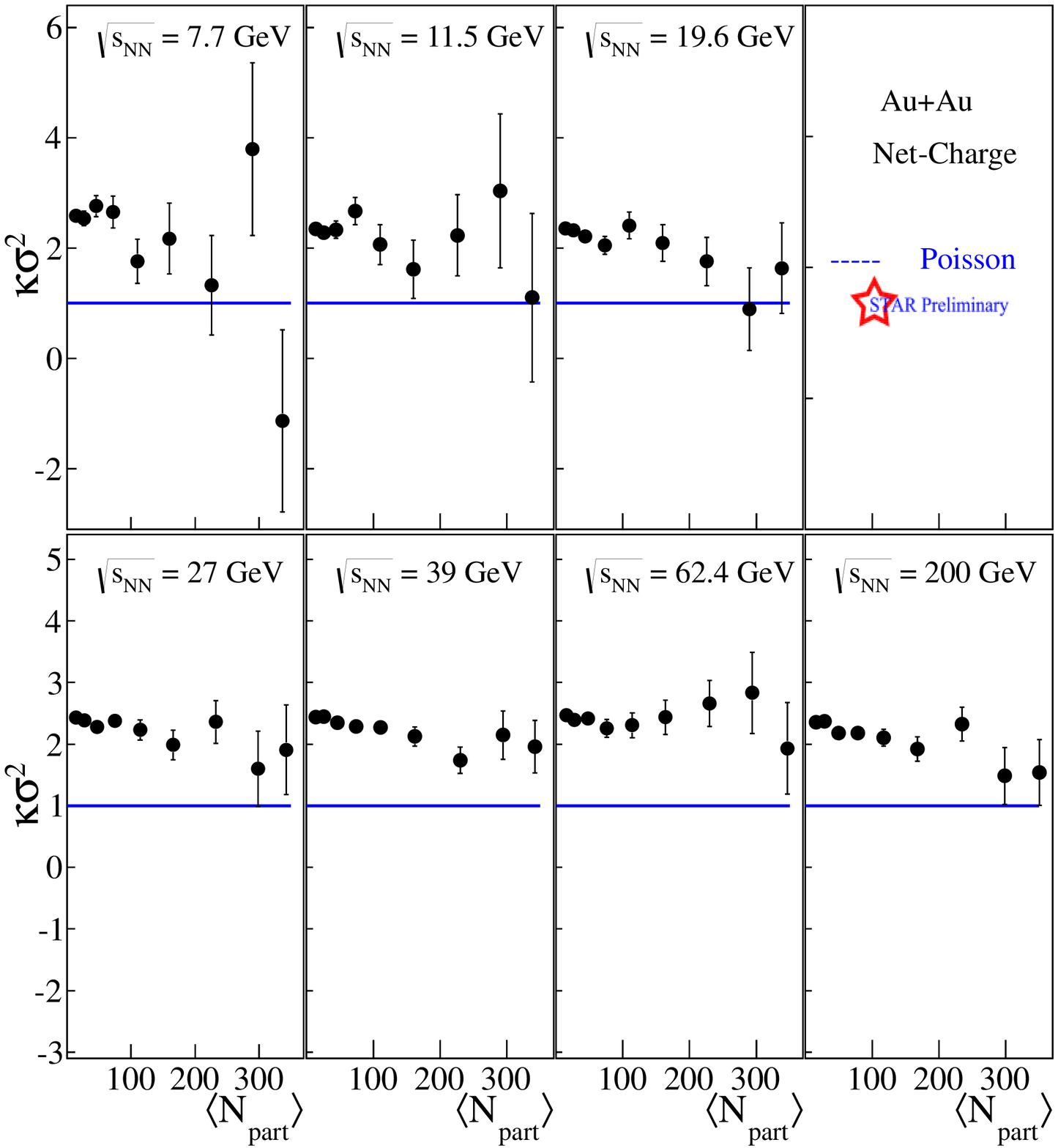}\caption{\label{fig:chargeksigsq}Centrality dependence of $\kappa$$\sigma^{2}$
of net-charge at $\sqrt{{\rm s}_{\rm NN}}$ = 7.7, 11.5, 19.6, 27,
39, 62.4, and 200 GeV. Shown in blue is the Poisson expectation.}
\end{minipage}\,\,\,\,\,\,%
\begin{minipage}[t]{0.5\columnwidth}%
\includegraphics[width=1\columnwidth]{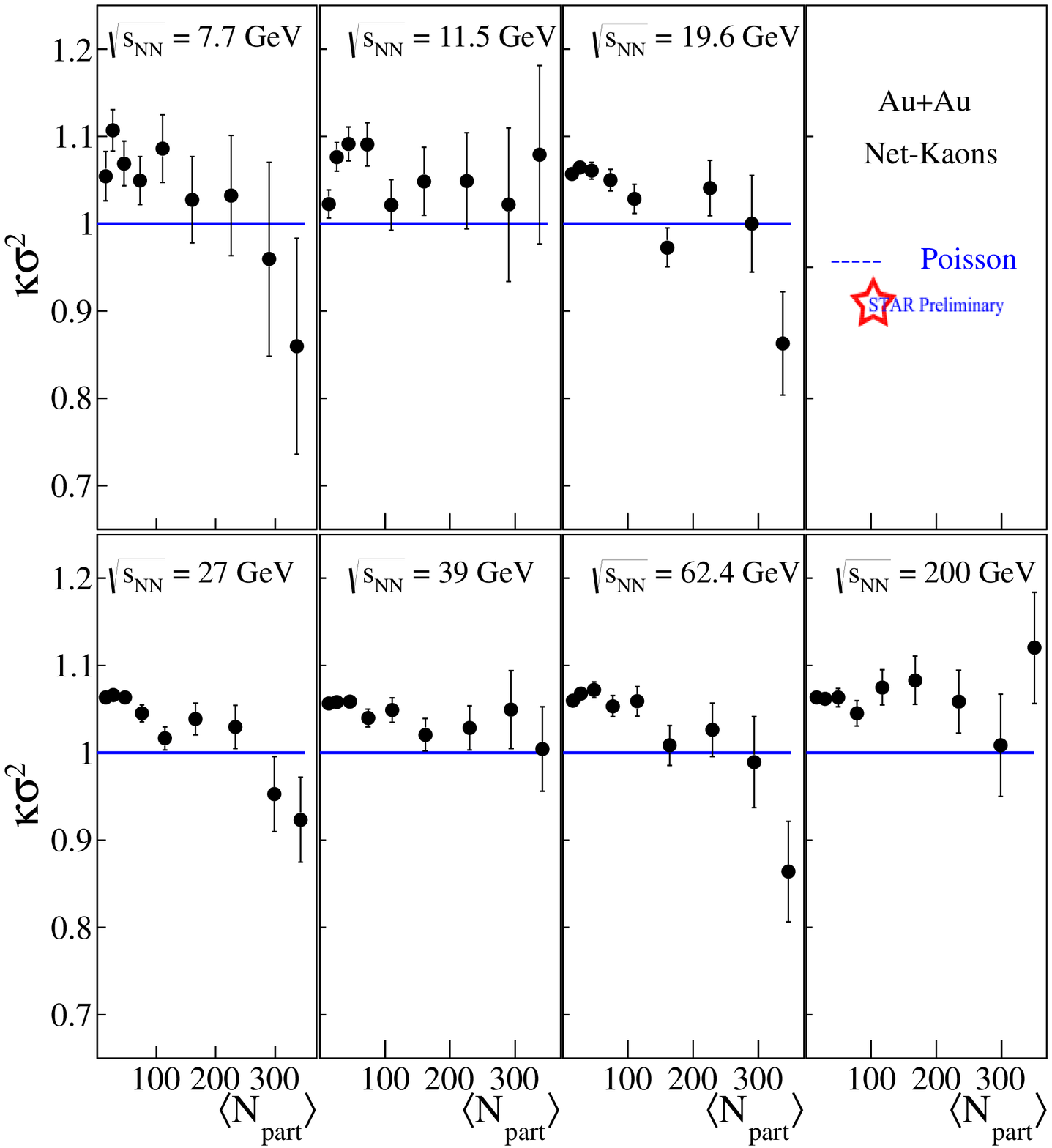}\caption{\label{fig:Kaonksigsq}Centrality dependence of $\kappa$$\sigma^{2}$
of net-kaons at $\sqrt{{\rm s}_{\rm NN}}$ = 7.7, 11.5, 19.6, 27,
39, 62.4, and 200 GeV. Shown in blue is the Poisson expectation.}
\end{minipage}
\end{figure}

The error bars shown are statistical only \citep{Luo12}. The moments
products were measured for net-charge and net-kaons ($K^{+}$$-$$K^{-}$)
and compared with the expectations from Poisson statistics and the
Hadron Resonance Gas (HRG) model \citep{KR11,Andro06}. In Poisson
statistics, the moments are functions only of the mean values (M):
$S\sigma=\nicefrac{(M^{+}-M^{-})}{(M^{+}+M^{-})}$ and $\kappa\sigma^{2}=1$.
In the HRG model, the moments are functions of thermodynamic parameters:
$S\sigma=\tanh(Q_{i}\mu_{Q}/T)$ and $\kappa\sigma^{2}\sim1.8$, where
$Q_{i}$$=$electric charge.

Shown in Fig.\ref{fig:chargeksigsq} and Fig.\ref{fig:Kaonksigsq}
is the centrality dependence of the fourth moment product $\kappa$$\sigma^{2}$
of net-charge and net-kaons, respectively, for seven beam energies
ranging from 7.7-200 GeV. Shown in blue is the Poisson expectation,
which is mathematically calculable using the mean values alone. The
moments product $\kappa$$\sigma^{2}$ is generally independent of
centrality and above the Poisson expectation at all beam energies.
The same general trends are also observed for the third moments product
$S\sigma$ for both net-kaons and net-charge.

\begin{figure}
\raggedright{}%
\begin{minipage}[t]{0.5\columnwidth}%
\includegraphics[width=1\columnwidth]{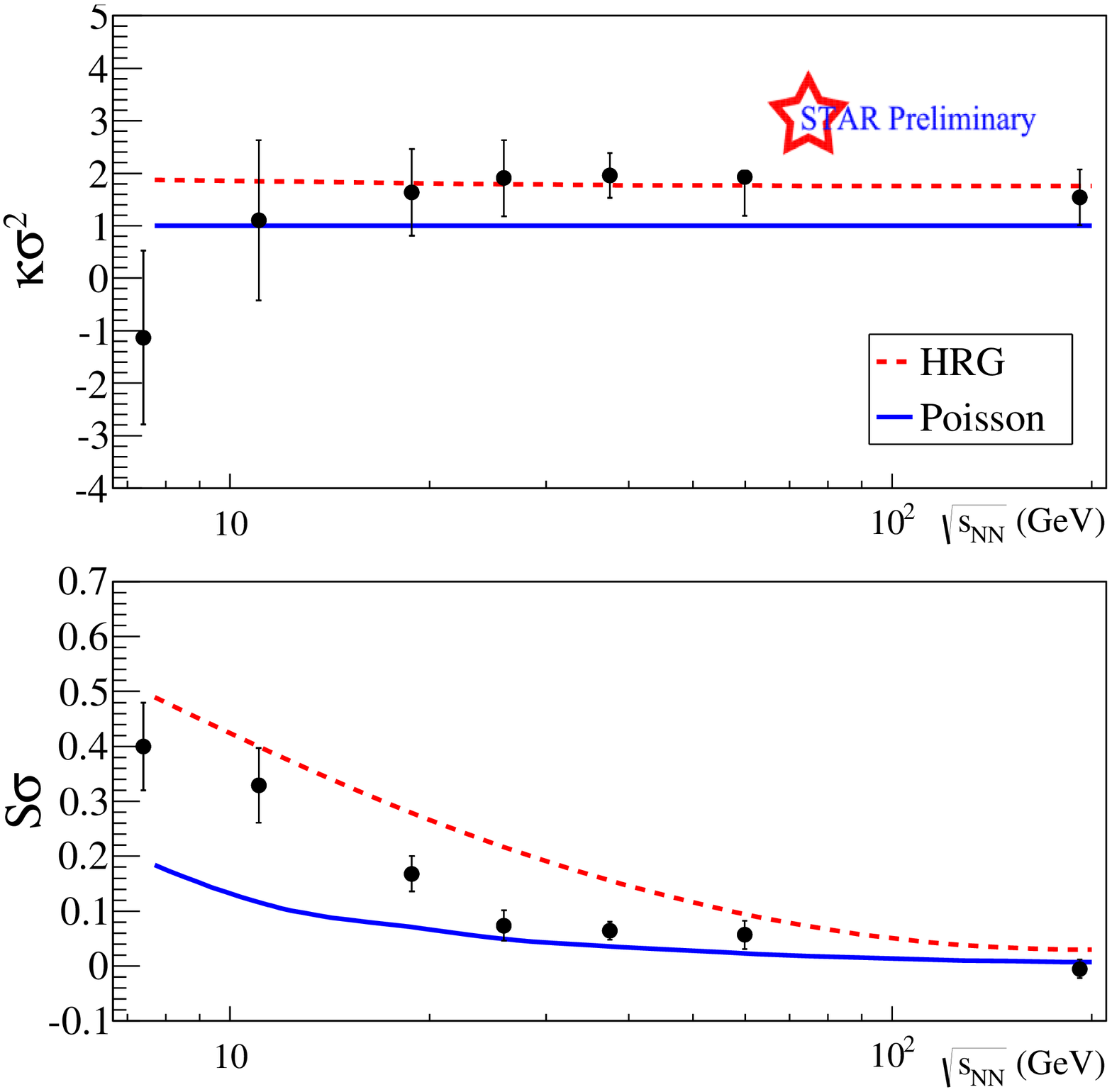}

\caption{\label{fig:charge05}Moments products S$\sigma$ and $\kappa$$\sigma^{2}$
for net-charge for 0-5\% central collisions. Solid line is the Poisson
expectation, and dashed line is the HRG expectation.}
\end{minipage}\,\,\,\,\,\,%
\begin{minipage}[t]{0.5\columnwidth}%
\includegraphics[width=1\columnwidth]{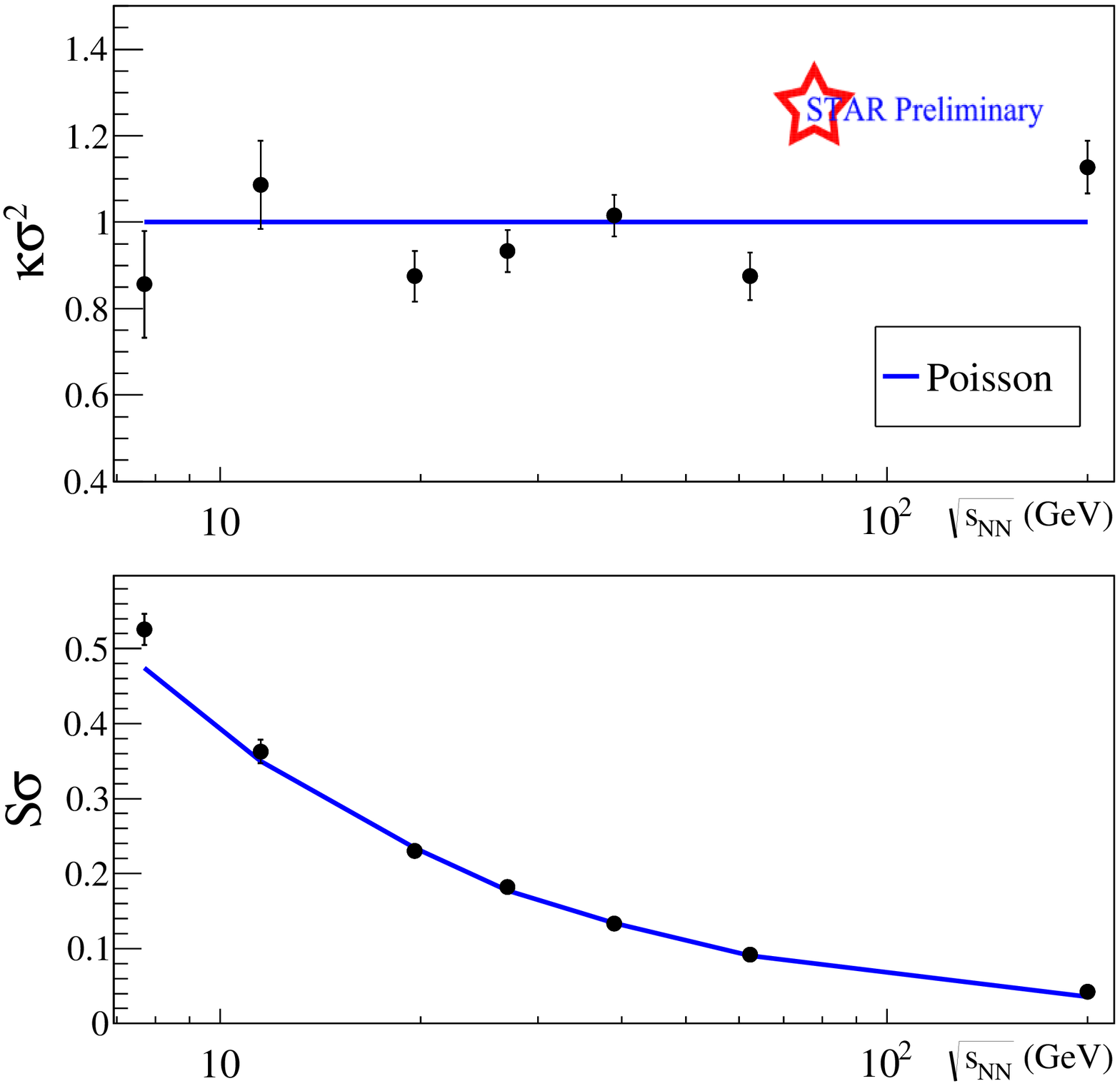}\caption{\label{fig:kaon05}Moments products S$\sigma$ and $\kappa$$\sigma^{2}$
for net-kaons for 0-5\% central collisions, with the Poisson expectation
shown as a solid line.}
\end{minipage}
\end{figure}

Shown in Fig.\ref{fig:charge05} is $\kappa$$\sigma^{2}$ (upper
frame) and $S\sigma$ (lower frame) for net-charge for 0-5\% central
collisions, where the solid line is the Poisson expectation and the
dashed line is the HRG expectation. $S\sigma$ is greater than the
Poisson expectation but less than the HRG expectation in central collisions.
No significant and $\sqrt{\rm s_{\rm NN}}$-localized enhancement
of either moments product is observed.

Shown in Fig.\ref{fig:kaon05} is $\kappa$$\sigma^{2}$ (upper frame)
and S$\sigma$ (lower frame) for net-kaons for 0-5\% central collisions,
where the solid line is the Poisson expectation. In general, the Poisson
expectation describes the moments products well.

\section{Summary}

STAR has measured S$\sigma$ and $\kappa$$\sigma^{2}$ of net-charge
and net-kaons at $\sqrt{{\rm s}_{\rm NN}}$ = 7.7, 11.5, 19.6, 27,
39, 62.4, and 200 GeV. The results are compared with Poisson and HRG
expectations. The moments products do not depend significantly on
centrality. For net-charge and net-kaons, the moments products are
generally above the Poisson expectation at all centralities and beam
energies. At the presently available beam energies, there is no beam-energy
and centrality localized large enhancements of the net-charge or net-kaons
moments products.

\section*{}

\bibliographystyle{model1-num-names}
\bibliography{danielbib}

\end{document}